\documentclass{article}

\def\mt{ L$_\mu -$ L$_\tau$ }
\usepackage{amsmath}	

\usepackage{amssymb}

\begin{document}

\begin{center}
Aiming for Unification of L$_\mu$-L$_\tau$ 
and the standard model gauge group \\
\vspace{1cm}
{Joe Sato
\footnote{Email:sato-joe-mc@ynu.ac.jp}
}\\
\vspace{1cm}
{
Department of Physics, Graduate School of Engineering Science,
\\ Yokohama National University, Yokohama, 240-8501, Japan
 }
\end{center}

%



\begin{abstract}
  In this letter we show a kind of \mt gauge symmetry can be unified
  into a simple group E$_7$ with the standard model gauge symmetry in the
  context of coset space unification. We also discuss some implication
  from one of this kind of unification.
\end{abstract}








\section{Introduction}
\label{Introduction}

The standard model (SM) is a very successful theory and well
established.  Indeed almost all terrestrial experiments are explained
by SM.  However, there are several questions: how to explain the
lepton flavor violation appearing in neutrino
oscillation~\cite{Fukuda_1998}, the existence of dark matter, the
baryon asymmetry, the discrepancy of muon anomalous magnetic moment
between the theoretical~\cite{Aoyama:2020ynm} and experimental
value~\cite{Abi:2021gix}, etc. To explain (some of) them many trials
have been made by extending the SM. As one of the direction in these
years many physicists extend the gauge symmetry,
say,\mt~\cite{Foot:1990mn,He:1990pn,Foot:1994vd,He:1991qd}.  It gives
plausible explanation~\cite{Gninenko:2001hx,Baek:2001kca,Ma:2001md}
for muon anomalous magnetic moment and, in addition, may give a
solution to Hubble inconsistency~\cite{Escudero:2019gzq,Araki:2021xdk}
and IceCube Gap too~\cite{Araki:2014ona,Kamada:2015era, Araki:2015mya,
  DiFranzo:2015qea}.


Besides those phenomenological questions, there are fundamental 
and/or conceptual questions in SM: Why are there three gauge groups
SU(3), SU(2), U(1)? Why does nature has matters, in other words,
for example why do quarks behave as a triplet of SU(3)?
Why are there three copies of materials? Why does exist three generations? 
The former can be explained partly by the unification
of the gauge group, the Grand Unified Theory (GUT). 


Then it is natural to ask whether \mt can be unified with the standard
model gauge group, i.e. grander unified theory. Naively it looks
difficult since (i) only leptons have \mt charge and (ii) there is a
generation dependence. In unified theories quarks are unified into
same multiplet with leptons and hence not only leptons but also quarks
should have \mt charge. Therefore we have to give up a simple (only
leptophilic) \mt and we also assign its charge to quarks.  On the
contrary generation dependence means that \mt is gauged family
symmetry.  Implementing them appropriately, in this letter we will see
a kind of \mt and SM gauge group are unified into a simple group
E$_7$ within the context of
coset space unification~\cite{Sato:1997hv}, a supersymmetric extension of nonlinear sigma model.

In Sec.\ref{CosetReview} we give a short review of coset space unification.
Then In Sec.\ref{Body} we show candidate assignments of \mt and
their interpretation. Finally we give a summary and discussion in sec.\ref{summary}.

\section{Coset space unification}
\label{CosetReview}
We first review the structure of coset space unification.  Three
family fermions including right-handed neutrinos naturally are
accommodated in the coset-space family
unification\cite{Buchmuller:1983iu} in supersymmetric (SUSY)
GUTs. Coset-spaces based on E$_7$ are known as unique choices to
contain three families of quarks and leptons~\cite{Kugo:1983ai}.  Among
them E$_7/$SU(5)$\times$U(1)$^3$ is the most interesting, since it
contains also three families of right-handed neutrinos as
Nambu-Goldstone (NG) multiplets~\cite{Yanagida:1985jc}. This model contains three families of
${\bf 10}_i + {\bf 5}^*_i + {\bf 1}_i$ ($i=1,2,3$) as NG multiplets.
Though in addition, there is an extra ${\bf 5}$, we ignore it in this
letter.  Here, the SU(5) is the usual GUT gauge group.  Their quantum
numbers under the unbroken subgroup are given in Table
\ref{charge_old}. Incidentally, though there is an extra ${\bf 5}$, we
will ignore it in this letter hereafter as it is irrelevant.
\begin{table}[ht]
\begin{center}
\begin{tabular}{c|ccc}
SU(5)
&\makebox[2cm]{U(1)$_1$}&\makebox[2cm]{U(1)$_2$}&\makebox[2cm]{U(1)$_3$}\\
\hline
{\bf 10}$_1$&0&0&4\\
{\bf 10}$_2$&0&3&-1\\
{\bf 10}$_3$&2&-1&-1\\
{\bf 5}$^*_1$&0&3&3\\
{\bf 5}$^*_2$&2&-1&3\\
{\bf 5}$^*_3$&2&2&-2\\
{\bf 1}$_1$&0&3&-5\\
{\bf 1}$_2$&2&-1&-5\\
{\bf 1}$_3$&2&-4&0\\
\end{tabular}
\caption{U(1) charges of the NG multiplets. The U(1)$_1$,
U(1)$_2$ and U(1)$_3$ are the unbroken U(1)'s of
coset-subspaces E$_7$/E$_6\times$U(1),
E$_6$/SO(10)$\times$U(1) and SO(10)/SU(5)$\times$U(1), respectively.}
\label{charge_old}
\end{center}
\end{table}
These U(1)'s are interpreted as those from the breaking chain
\begin{eqnarray}
 {\rm E}_7 &\longrightarrow&  {\rm E}_6\times {\rm U(1)}_1
\longrightarrow
{\rm SO(10)}\times {\rm U(1)}_1\times{\rm U(1)}_2
\nonumber
\\& \longrightarrow &
{\rm SU(5)}\times{\rm U(1)}_1\times{\rm U(1)}_2 \times{\rm U(1)}_3
\label{breaking}
\end{eqnarray}
The subscripts of fields in table~\ref{charge_old} (and also following
tables) are not generation indices but the embedding (or definition)
of fields into E$_7$ adjoint representation. For example at the first
breaking, {\bf 1}$_3$, {\bf 1}$_2$, {\bf 5}$^*_3$, {\bf 5}$^*_2$, and {\bf
  10}$_3$ appear in the spectrum. Similarly {\bf 1}$_1$ {\bf 5}$^*_1$
{\bf 10}$_2$ does at the second breaking.  At the last {\bf 10}$_1$
arises.

\section{ Rearrangement of U(1) charge for \mt}
\label{Body}
As U(1)'s are commutable and hence we can take linear combination,
that is,
\begin{equation}
Q_i = a_{ij}q_j,
\end{equation}
where $q_j$ is given in the table \ref{charge_old} and $Q_i$'s are the
new U(1) charges.  The existence of \mt indicates that one of
rearranged U(1), say new U(1)$_3$, charge $Q_3$ for a pair of {\bf
  10} and {\bf 5} must be $\pm 1, 0$.  Indeed there are three kinds of
such recombination, given by
\begin{eqnarray}
 a_{ij}&=&
\begin{pmatrix}
1&0&0\\
0&-\frac{5}{4}&-\frac{3}{4}\\
 0&-\frac{1}{4}&\frac{1}{4}
\end{pmatrix}
\label{deform1}\\
&=&
\begin{pmatrix}
\frac{2}{3}&\frac{1}{12}&\frac{1}{4}\\
0&\frac{5}{4}&-\frac{3}{4}\\
 -\frac{1}{3}&\frac{1}{12}&\frac{1}{4}
\end{pmatrix}
\label{deform2}\\
&=&
\begin{pmatrix}
\frac{2}{3}&\frac{1}{3}&1\\
-\frac{5}{3}&-\frac{5}{6}&-\frac{1}{2}\\
-\frac{1}{3}&\frac{1}{3}&0
\end{pmatrix}.
\label{deform3}
\end{eqnarray}
 Each solution gives independent breaking chain.

\begin{table}[ht]
\begin{center}
\begin{tabular}{c|ccc}
SU(5)
&\makebox[2cm]{U(1)$_1$}&\makebox[2cm]{U(1)$_2$}&\makebox[2cm]{U(1)$_{3=\mu-\tau}$}\\
\hline
{\bf 10}$_1$&0&3&1\\
{\bf 10}$_2$&0&3&-1\\
{\bf 10}$_3$&2&-2&0\\
{\bf 5}$^*_1$&0&6&0\\
{\bf 5}$^*_2$&2&1&1\\
{\bf 5}$^*_3$&2&1&-1\\
{\bf 1}$_1$&0&0&-2\\
{\bf 1}$_2$&2&-5&-1\\
{\bf 1}$_3$&2&-5&1\\
\end{tabular}
\caption{U(1) charges of the NG multiplets in breaking
  (\ref{deform1}).  The U(1)$_1$, U(1)$_2$ and U(1)$_3$ are the
  unbroken U(1)'s of coset-subspaces E$_7$/E$_6\times$U(1),
  E$_6$/SU(5)$\times$SU(2)$\times$U(1) and SU(2)/U(1), respectively.}
\label{charge_new1}
\end{center}
\end{table}
The first recombination leads new U(1) charges given 
in table~\ref{charge_new1}.  These U(1)'s are interpreted as residual one from
the breaking chain
\begin{eqnarray}
 {\rm E}_7 &\longrightarrow&  {\rm E}_6\times {\rm U(1)}_1
\longrightarrow
{\rm SU(5)}\times {\rm SU(2)}\times {\rm U(1)}_1\times{\rm U(1)}_2
\nonumber
\\& \longrightarrow &
{\rm SU(5)}\times{\rm U(1)}_1\times{\rm U(1)}_2 \times{\rm U(1)}_{3=\mu-\tau}
\label{breaking1}
\end{eqnarray}
In this chain both ({\bf 5}$^*_3$ , {\bf 5}$^*_2$)
and ({\bf 10}$_2$,{\bf 10}$_1$) appear as SU(2) doublet at the second breaking.

\begin{table}[ht]
\begin{center}
\begin{tabular}{c|ccc}
SU(5)
&\makebox[2cm]{U(1)$_1$}&\makebox[2cm]{U(1)$_2$}&\makebox[2cm]{U(1)$_{3=\mu-\tau}$}\\
\hline
{\bf 10}$_1$&1&-1&1\\
{\bf 10}$_2$&0&4&0\\
{\bf 10}$_3$&1&-1&-1\\
{\bf 5}$^*_1$&1&3&1\\
{\bf 5}$^*_2$&2&-2&0\\
{\bf 5}$^*_3$&1&3&-1\\
{\bf 1}$_1$&1&-5&1\\
{\bf 1}$_2$&0&0&-2\\
{\bf 1}$_3$&1&-5&-1\\
\end{tabular}
\caption{U(1) charges of the NG multiplets in breaking
  (\ref{deform2}).  The U(1)$_1$, U(1)$_2$ and U(1)$_3$ are the
  unbroken U(1)'s of coset-subspaces
  E$_7$/SO(10)$\times$SU(2)$\times$U(1), SO(10)/SU(5)$\times$U(1) and
  SU(2)/U(1), respectively.}
\label{charge_new2}
\end{center}
\end{table}
The second one arises from the
breaking chain
\begin{eqnarray}
 {\rm E}_7 &\longrightarrow&  {\rm SO(10)}\times {\rm SU(2)}\times
{\rm U(1)}_1\nonumber\\
&\longrightarrow&
{\rm SU(5)}\times {\rm SU(2)}\times {\rm U(1)}_1\times{\rm U(1)}_2
\nonumber
\\& \longrightarrow &
{\rm SU(5)}\times{\rm U(1)}_1\times{\rm U(1)}_2 \times{\rm U(1)}_{3=\mu-\tau}
\label{breaking2}
\end{eqnarray}
Their U(1) charges are shown in table~\ref{charge_new2}.  In this
chain both SO(10) {\bf 16}=({\bf 10}$_1$+{\bf 5}$^*_1$+{\bf 1}$_1$)
and ({\bf 10}$_3$+{\bf 5}$^*_3$+{\bf 1}$_3$) form an SU(2) doublet at
the first breaking.  Note that U(1) charges for {\bf 1}$_1$ are
reversed from the naive change by (\ref{deform2}). It is due to the
fact that we always have a freedom of choice to extracting a
representation {\bf r} or {\bf r}$^*$ as NG boson.  As we need GUT
representation while we have no choice to select say, {\bf 10}$^*$, we
can switch {\bf 1} to {\bf 1}$^*$. It may lead drastic change of
phenomenology though we will not touch this point in this letter.

\begin{table}[ht]
\begin{center}
\begin{tabular}{c|ccc}
SU(5)
&\makebox[2cm]{U(1)$_1$}&\makebox[2cm]{U(1)$_2$}&\makebox[2cm]{U(1)$_{3=\mu-\tau}$}\\
\hline
{\bf 10}$_1$&4&2&0\\
{\bf 10}$_2$&0&-3&1\\
{\bf 10}$_3$&0&-3&-1\\
{\bf 5}$^*_1$&4&-1&1\\
{\bf 5}$^*_2$&4&-1&-1\\
{\bf 5}$^*_3$&0&-6&0\\
{\bf 1}$_1$&-4&-5&-1\\
{\bf 1}$_2$&-4&-5&1\\
{\bf 1}$_3$&0&0&-2\\
\end{tabular}
\caption{U(1) charges of the NG multiplets in breaking
  (\ref{deform3}).  The U(1)$_1$, U(1)$_2$ and U(1)$_3$ are the
  unbroken U(1)'s of coset-subspaces
  E$_7$/SU(6)$\times$SU(2)$\times$U(1), SU(6)/SU(5)$\times$SU(1) and
  SU(2)/U(1), respectively.}
\label{charge_new3}
\end{center}
\end{table}
The final one corresponds to the breaking chain
\begin{eqnarray}
 {\rm E}_7 &\longrightarrow& ({\rm SU(6)}\times {\rm SU(2)}\times {\rm U(1)}_1)\nonumber\\
&\longrightarrow&
{\rm SU(5)}\times {\rm SU(2)}\times {\rm U(1)}_1\times{\rm U(1)}_2
\nonumber
\\& \longrightarrow &
{\rm SU(5)}\times{\rm U(1)}_1\times{\rm U(1)}_2 \times{\rm U(1)}_{3=\mu-\tau}
\label{breaking3}
\end{eqnarray}
In this chain both ({\bf 5}$^*_1$,{\bf 5}$^*_2$) and ({\bf
  10}$_2$,{\bf 10}$_3$) appear as SU(2) doublet at the second
breaking.

Again U(1) charges for {\bf 1}$_1$ and {\bf 1}$_2$ are reversed from
the naive change by (\ref{deform3}). In addition in this breaking
chain we should not have the first stage, that is, we should interpret
that E$_7$ breaks directly to SU(5)$\times$SU(2)$\times$U(1). Otherwise
we could not realize three {\bf 10}'s.  These lead drastic change of
phenomenology too though we will not touch this point in this letter. 

Thus there are essentially these three chains. It is understood by
following two steps.  The first one is that there are four maximal
subgroups of E$_7$ including SU(5), which are E$_6\times$U(1),
SO(12)$\times$SU(2), SU(8), and SU(6)$\times$SU(3).  The second is
among them we can directly check that it is impossible to get three
{\bf 10} and three {\bf 5}$^*$ via SU(8) by direct calculation.  We
note also that these three chains are independent. It is understood by
the fact that U(1) charges for ``right-handed neutrinos" are
different. Therefore in each chain, in principle, we will have a quite
different phenomenology for, at least, neutrino physics.

Incidentally, new U(1)$_1$ and U(1)$_2$ for the first breaking chain
while $q_2$ and $q_3$ are exchangeable in the second and the third
chain. This exchange may lead to different phenomenology.

\paragraph{Matter assignment -- examples}
To discuss phenomenology it is necessary to specify an assignment of 
fermions. To do so we need to know the breaking 
parameter~\cite{Sato:1997hv}. While
the first breaking chain looks similar to the previous model the others
look quite different because the origin of the right-handed neutrino(s)
is different. Therefore, we show examples from the breaking chain
(\ref{breaking1}).  In this U(1) assignment, $\mu$-flavored doublet
belongs to {\bf 5}$^*_2$ and $\tau$-flavored one does {\bf
  5}$^*_3$. Correspondingly $\mu$-flavored singlet belongs to {\bf
  10}$_2$ and $\tau$-flavored one does belongs to {\bf 10}$_1$. The remaining
$e$-flavord leptons are contained in {\bf 5}$^*_1$ and {\bf 10}$_3$.
The embedding of other fermions is arbitrary. Though it is determined by the
mass spectrum of fermions.
To do so we need breaking parameters but 
it is totally beyond the scope of this letter. 
Instead of specifying breaking parameters, 
we show two possible examples of the emmbeddings as examples.

The first one is
\begin{eqnarray}
{\bf 10}_1=(t^c,\{t_L,b_L\},\tau^c)&\ 
&{\bf 5}^*_1=(d^c,\{\nu_{Le}, e_L\}),
\\
{\bf 10}_2=(c^c,\{c_L,s_L\},\mu^c)&\ 
&{\bf 5}^*_2=(s^c,\{\nu_{L\mu}, \mu_L\}),
\\
{\bf 10}_3=(u^c,\{u_L,d_L\},e^c)&\ 
&{\bf 5}^*_3=(b^c,\{\nu_{L\tau}, \tau_L\}).
 \end{eqnarray}
This keeps the naive structure of generation though from the
breaking pattern it may be difficult to assign the 1st(3rd)
generation into ${\bf 10}_{3(1)}$. 

With this assignment, the coupling of fermions with \mt gauge boson
$Z'$ is given by
\begin{eqnarray}
 \mathcal{L}_{Z'}=g_{Z'}
\{
(\bar\mu\gamma^\rho\mu+\bar\nu_{L\mu}\gamma^\rho\nu_{L\mu})
-(\bar\tau\gamma^\rho\tau+\bar\nu_{L\tau}\gamma^\rho\nu_{L\tau})
\nonumber
\\ +(\bar c\gamma^\rho\gamma_5 c-\bar s\gamma^\rho s)
-(\bar t\gamma^\rho\gamma_5 t-\bar b\gamma^\rho b)
\}Z'_\rho .
\end{eqnarray}
Leptons have a vector coupling with \mt gauge boson appropriately.
On the contrary quarks have
an axial vector coupling for $c$ and $t$ while a vector coupling for
$s$ and $b$. At this moment there is no strong constraint from
experiments.


The second one is
\begin{eqnarray}
{\bf 10}_1=(u^c,\{u_L,d_L\},\tau^c)&\ 
&{\bf 5}^*_1=(b^c,\{\nu_{Le}, e_L\}),
\\
{\bf 10}_2=(c^c,\{c_L,s_L\},\mu^c)&\ 
&{\bf 5}^*_2=(d^c,\{\nu_{L\mu}, \mu_L\}),
\\
{\bf 10}_3=(t^c,\{t_L,b_L\},e^c)&\ 
&{\bf 5}^*_3=(s^c,\{\nu_{L\tau}, \tau_L\}).
 \end{eqnarray}
This emmbeddings is achieved by assuming that 
the prediction for quark mixing given 
in~\cite{Sato:1997hv} holds even after the recombination. 
By this the
embedding for quark doublets is also determined. We still have the
freedom for the embedding of right-handed quarks. Here we examine one of
the possibilities that look most interesting.

With this assignment, the coupling of fermions with \mt gauge boson
$Z'$ is given by
\begin{eqnarray}
 \mathcal{L}_{Z'}=g_{Z'}
\{
(\bar\mu\gamma^\rho\mu+\bar\nu_{L\mu}\gamma^\rho\nu_{L\mu})
-(\bar\tau\gamma^\rho\tau+\bar\nu_{L\tau}\gamma^\rho\nu_{L\tau})
\nonumber
\\ -(\bar u\gamma^\rho\gamma_5u+\bar d\gamma^\rho\gamma_5d)
+(\bar c\gamma^\rho\gamma_5 c+\bar s\gamma^\rho\gamma_5s)
\}Z'_\rho .
\end{eqnarray}
Again leptons have a vector coupling with \mt gauge boson appropreately.
On the contrary quarks have
an axial vector coupling with it.
Therefore in the non-relativistic limit there is no
connection between quarks and leptons mediated by the \mt gauge boson.
 For example there
is no constraint from atomic physics.

There may be an effect on meson decay. However, as there is no direct
coupling of $Z'$ to electrons, very tiny effects are expected and
hence we would expect that this model is also free from constraints on
mesons.

Another implication is on proton decay. Though for dimension-5
operators we have no indication, for dimension-6 operator that is that
mediated by gauge bosons we have an interesting ``prediction". Proton
decay is induced by
$\overline{\bf 10}_1{\bf 10}_1\overline{\bf 5}^*_2{\bf 5}^*_2$. It leads
\begin{equation}
    p\rightarrow \mu^+\pi^0 .
\end{equation}
Instead of $e^+$ we will observe $\mu^+$. If we find this then it is an prominent signature of the scenario.

\section{Summary and Discussion}
\label{summary}
In this pape, we show a unification of SM gauge and \mt gauge
symmetry into the simple group E$_7$ in the context of coset space
unification. There are three types of unification that will lead
to different phenomenology, at least for neutrino. To check it we need to
specify breaking chains and breaking parameters as 
in~\cite{Sato:1997hv}. 

  Even though details are a matter of breaking parameters, we show two
examples of matter assignment for the first breaking chain 
to show that this framework contains a plenty of models. Indeed derived
low energy Lagrangians are quite different from each other and possible
predictions are distinctive.

In addition to those breaking parameters for matter assignment, we have
to specify the breaking mechanism to seek the final theory.  Mechanism
is strongly related with not only what kind of matter appear in the
spectrum but also breaking parameters which determine low energy
Lagrangian, say yukawa terms.  There are several breaking methods:
Spontaneous symmetry breaking, Coset Space Dimensional
Reduction~\cite{Manton:1979kb,Kapetanakis:1992hf}, Difference of
boundary condition between bosons and
fermions~\cite{Scherk:1978ta,Scherk:1979zr}, Non-linear
realization~\cite{Buchmuller:1983iu,Bando:1983ab,Bando:1984cc,Bando:1984fn,Higashijima:1999ki}, Hosotani mechanism~\cite{Hosotani:1983xw}, 
etc. 

%

Some of those mechanisms requires gauge
symmetry\cite{Manton:1979kb,Kapetanakis:1992hf,Higashijima:1999ki}. Note
that the coset space unification is a kind of non-linear realization and
hence only the global symmetry is relevant. To construct a full gauge
theory we have to ensure that the global symmetry can be gauged. Indeed
these breaking mechanisms rely on the fact that the fundamental symmetry,
which is G of coset G/H, is gauged. By combining with these mechanism
we will find a way to gauge SU(5).

All of the details are beyond the scope of this
work, thus it will be made in the future. 

\vspace{1mm}
\paragraph{Acknowledgment}
The author thanks MIZOGUCHI, Shun'ya for helpful implication during the workshop ``Particle and phenomenology 
workshop 2020" held in the Nambu Yoichiro Institute of Theoretical 
and Experimental Physics (NITEP) at Osaka City University.
The author is also grateful to TAKANISHI, Yasutaka for useful comment.
This work was supported by JSPS KAKENHI Grants No. JP18H01210 and
MEXT KAKENHI Grant No. JP18H05543.

\bibliographystyle{unsrt}
\bibliography{unif}





\end{document}